\documentclass{aa}
\usepackage{graphicx}

\newcommand\simlt{\lower.5ex\hbox{$\; \buildrel < \over \sim \;$}}
\newcommand\simgt{\lower.5ex\hbox{$\; \buildrel > \over \sim \;$}}

\newcommand\be{\begin{equation}}
\newcommand\ee{\end{equation}}
\newcommand\ba{\begin{eqnarray}}
\newcommand\ea{\end{eqnarray}}

\begin{document}

\title{Magnetic Rayleigh-Taylor instability\\
for Pulsar Wind Nebulae in expanding Supernova Remnants}

\author{ 
N. Bucciantini \inst{1,2}, E. Amato \inst{3}, R. Bandiera\inst{3}, J.~M. Blondin\inst{2}, L. Del Zanna \inst{1}}

\offprints{N.Bucciantini\\
 e-mail: niccolo@arcetri.astro.it}

\institute{Dip. di Astronomia e Scienza dello Spazio,
  Universit\`a di Firenze, Largo E.Fermi 2, I-50125 Firenze, Italy\\
\and
Department of Physics, North Carolina State University,
Raleigh, NC 27695, USA\\
\and
INAF, Osservatorio Astrofisico di Arcetri,
 Largo E.Fermi 5, I-50125 Firenze, Italy
}
\authorrunning{N.Bucciantini et al.}
\titlerunning{Magnetic Rayleigh-Taylor instability in PWN-SNR systems}
\date{Received  1 March 2004 / Accepted 5 May 2004}

\abstract{We present a numerical investigation of the development of
Rayleigh-Taylor instability at the interface between an expanding Pulsar 
Wind Nebula and its surrounding Supernova Remnant. These systems have long
been thought to be naturally subject to this kind of instability, given
their expansion behavior and the density jump at the contact discontinuity.
High resolution images of the Crab Nebula at optical frequencies show 
the presence of a complex network of line-emitting filaments protruding 
inside the synchrotron nebula. These structures are interpreted as the 
observational evidence that Rayleigh-Taylor instability is in fact at work. 
The development of this instability in the regime appropriate to
describe Supernova Remnant-Pulsar Wind Nebula systems is non-trivial.
The conditions at the interface are likely close to the stability threshold,
and the inclusion of the nebular magnetic field, which might play an 
important role in stabilizing the system, is essential to the modeling. 
If Rayleigh-Taylor features can grow efficiently a mixing layer
in the outer portion of the nebula might form where most of the supernova
material is confined. When a magnetic field close to equipartition is
included we find that the interface is stable, and that even a weaker
magnetic field affects substantially the growth and shape of the fingers.
\keywords{Instabilities -- MHD -- pulsars: general -- 
stars: outflows -- ISM: supernova remnants}
}
\maketitle

\section{Introduction}

Pulsars are rapidly rotating magnetized neutron stars that usually form as 
the result of a Supernova (SN) explosion. As a consequence of the 
electromagnetic 
torques acting on it, a pulsar releases most of its rotational energy in the 
form of a relativistic magnetized wind. The wind is usually thought to be made 
of electron-positron pairs and to carry a magnetic field that far enough 
from the 
light cylinder is almost purely toroidal (\cite{michel99}; \cite{goldreich69}). 
This outflow is highly relativistic, with a terminal Lorentz factor in the range 
$10^4$-$10^7$. Its confinement by the surrounding Supernova Remnant (SNR) 
generates a nebula of relativistically hot material that shines through 
synchrotron and Inverse Compton emission from radio wavelengths up to 
$\gamma$-rays: this is what we call a pulsar wind nebula (PWN) or ``plerion''.

During a SN explosion as much as $10^{51}$ erg of energy are released 
in the form of a blast wave that produces a strong shock propagating in the 
surrounding medium. The ejected material is initially heated by the blast 
wave and set into motion. As the ejecta expand, their thermal pressure 
finally becomes so low as to be dynamically unimportant: from this moment 
on the expansion can be approximated as homologous 
(\cite{chevalier89}; \cite{matzner99}). This phase is 
referred to as ``free expansion'' of the ejecta.

The evolution of the PWN inside the free expanding ejecta depends on many 
different parameters such as the pulsar luminosity, the flux anisotropies of the 
pulsar wind (\cite{komissarov03}; \cite{delzanna04}), the density and velocity 
distribution in the ejecta (\cite{dwarkadas98}; \cite{featherstone01}; 
\cite{blondin96}), as well as the presence of large and/or small scale 
anisotropies (\cite{chevalier89}; \cite{campbell03}). As a consequence, the 
detailed modeling of a single PWN-SNR system requires the knowledge of a 
number of parameters depending on the specific conditions, that are usually
unknown.

The simplest approximation one can make for the time evolution of the
PWN size is obtained assuming constant pulsar luminosity and spherical 
symmetry (\cite{chevalier92}; \cite{swaluw01}; 
\cite{bucciantini03}; \cite{bucciantini04}). If a radial power law density 
profile, $\rho \propto r^{-\alpha}t^{\alpha-3}$, is further assumed for the 
SN ejecta, then 
the PWN size evolves as $t^{(6-\alpha)/(5-\alpha)}$. For a more detailed 
description of the various phases of the PWN-SNR evolution see 
Bucciantini et al. (2003) and references therein.

The interface between the synchrotron nebula and the swept up shell of ejecta 
has been thought to be Rayleigh-Taylor (hereafter RT) unstable 
(\cite{chevalier75}; \cite{bandiera83}). In the case of the Crab Nebula the
RT instability is expected to be at the origin of the complex network of
emission-line 
filaments protruding into the PWN (\cite{hester96}, H96 hereafter). The recent 
images by H96 show that these ``radial'' filaments are joined at their basis 
by faint, thin , ``tangential'', and often somewhat arcuate features, which 
are interpreted as tracing the location of the RT unstable interface. 
The filamentary structure presents a clear hierarchy as expected from a 
multimode instability.

Simulations of the RT instability in the first phase of the PWN-SNR evolution 
have been presented by Jun in a classical hydrodynamical (HD) regime 
(\cite{jun98}, hereafter J98). However, as H96 pointed out, the standard 
model for PWNe (\cite{kennel84}) leads to believe that many of them are 
magnetically dominated 
at the contact discontinuity with the SNR. In addition, within the standard 
framework, the nebular magnetic field, usually expected to be purely toroidal, 
is tangential to the contact discontinuity. 
A parallel 
magnetic field is expected to have a stabilizing effect and possibly even  
to suppress the formation of fingers 
(\cite{chandrasekhar61}; \cite{wang83}; \cite{jun95}). In the recent work by H96 
a comparison with results from classical magnetohydrodynamical (MHD) 
simulations of the RT instability 
(\cite{jun95}) was carried out. The results suggest that, in the case 
of the Crab Nebula, the magnetic field should be close to the critical value for 
stability.
However, the conditions at the contact discontinuity between PWNe and SNRs 
are quite different from those adopted both in the formulation of the standard 
theory (\cite{chandrasekhar61}) and in the existent MHD simulations 
(\cite{jun95}), as we will discuss in the following. 

In this paper we present a study of the RT instability in the presence of 
a tangential magnetic field in the context of PWN-SNR systems. Our analysis
is carried out by means of 2D special relativistic MHD (RMHD) simulations. 
In Section 2 we review the standard theory for plane-parallel RT instability,
discussing how the main results are modified in the situation under investigation. 
In Section 3 the 
numerical method and the initial conditions for the simulations are described. 
Section 4 is dedicated to the numerical results both in the HD and MHD 
regimes. In Section 5 we finally summarize our conclusions.

\section{The Rayleigh-Taylor instability}

RT instability occurs when a heavy fluid is supported by a lighter fluid in a 
gravitational field, or, equivalently, when a heavy fluid is accelerated by a 
lighter fluid. This instability can be at work in many different astrophysical 
contests ranging from supernova explosions (\cite{fryxell91}) to shock wave 
interaction with ISM clouds (\cite{stone92}), from accretion onto compact 
objects (\cite{wang83}) to SNR evolution (\cite{chevalier92b}).

In the linear regime, short wavelengths grow faster than long wavelengths. When 
the system enters the non linear phase the fluid interface assumes its 
characteristic ``mushroom-finger'' structure, with the finger penetrating the 
lighter fluid. As RT fingers grow, the shear between the two fluids gives rise 
to secondary Kelvin-Helmholtz (KH hereafter) instabilities, which concentrate 
vorticity in a mushroom cap at the tip of the fingers. The cap increases the 
drag on the finger slowing down its growth. In the fully non linear phase the 
evolution of the instability is affected by effects such as mergers and
breakup of the fingers, which form a mixing layer between the two fluids.

The evolution of the RT instability is influenced by many different factors.
Viscosity tends to reduce the growth rate and to stabilize the system
(\cite{plesset74}). Compressibility (i.e.\ radiative cooling) can produce 
thinner and longer fingers (\cite{gardner88}). However, the most important
effects in the astrophysical context are probably those due to the presence 
of a magnetic field. In general the magnetic field will have both a normal 
and a tangential component to the 
interface. We will deal only with the effect of a tangential magnetic field. 
In fact, pulsar winds, and, as a consequence, PWNe 
(at least to first approximation) are expected to contain a purely
toroidal magnetic field. We will here neglect global instabilities that can remove 
the axial symmetry of the system, like kink instability (\cite{begelman98}), and
generate a non-negligible magnetic field component perpendicular to the interface. 
The RT instability at the contact discontinuity with 
the SNR can in principle generate a radial component as magnetic field lines 
are forced to bend along the finger.

Let us recall the results of the linear theory for the simple plane parallel 
case with a uniform tangential magnetic field in both fluids, 
under the assumption of incompressibility. This situation  
was first studied analytically by Chandrasekhar (1961). The linear stability 
theory shows that a tangential magnetic field slows the 
growth of the RT instability.
The growth rate for modes with wave number $k$ parallel to the magnetic 
field lines is given by:
\be
w^{2}=\frac{gk(\rho_{2}-\rho_{1})-B^{2}k^{2}/2\pi}{(\rho_{2}+\rho_{1})}\ ,
\label{eq:rtch}
\ee
where $\rho_{1}$ and $\rho_{2}$ are the densities of 
the light and heavy fluid respectively, $B$ is the magnetic field, 
$g$ is the gravitational acceleration (or equivalently the acceleration 
of the system) and the real part of $w$ is the growth rate.

From Eq.~\ref{eq:rtch} one derives the critical strength the tangential magnetic 
field should have to stabilize the system against perturbations of 
wavelength $\lambda$ or smaller:
\be
B_{c}=\sqrt{g\lambda(\rho_{2}-\rho_{1})}.
\label{eq:rtbc}
\ee
Similarly, the critical wavelength for a given magnetic field is defined as:
\be
\lambda_{c}=\frac{B^2}{g(\rho_{2}-\rho_{1})};
\label{eq:rtlc}
\ee
the instability will be suppressed for smaller scales.

In the work by H96 these 
equations were used to interpret the filamentary network observed in the Crab 
Nebula as the result of the RT instability in PWN-SNR systems. However, 
many of the assumptions of the plane parallel theory do not apply in the 
context of PWNe: the lighter fluid is relativistically hot, the density in 
the swept-up shell is not uniform, nor is the magnetic field (which is 
actually relevant only on one side of the interface, the PWN's), 
all quantities evolve in time so that 
no fixed background conditions can be assumed, and, finally, 
spherical geometry must be used. 

Concerning the relativistic corrections, Allen \& Hughes (1984) have shown 
that in the HD case the growth rate is still given by Eq.~\ref{eq:rtch}
substituting the energy density with the enthalpy. The latter is, for a 
relativistic magnetized fluid:
\be
h=\rho c^2 +{\Gamma \over {\Gamma -1}}P+{B^{2} \over 4 \pi},
\label{eq:relenth}
\ee
with $\rho$ the rest mass density, $P$ the thermal pressure, 
$B$ the magnetic field, and $\Gamma=4/3$ the adiabatic coefficient.

Corrections to the incompressible approximation (for modes parallel to 
the magnetic
field), in the case of a strong magnetic field, have been presented so far 
only to the first order in $B^{2}/P$ by Shivamoggi (1982), who showed that
Eq.~\ref{eq:rtch} holds but the denominator and the term in $B^{2}$  have 
to be multiplied by a factor $1+B^2/(8 \pi \Gamma P)$. 
The second order effect of a strong magnetic field appears to be that of 
further reducing the growth rate and increasing the stability of the system.
Effects due to magnetic pressure are probably the reason why Wang \& Nepveu 
(1983) 
in simulations of accretion onto compact objects with high magnetic fields 
found that (with respect to the standard stability value): 
``considerably weaker fields are sufficient to stop the infall''.

\subsection{Stability criterion and growth rate in the framework of
 the self-similar evolution}
\label{sect:stcrit}

The equations above are only valid in the plane parallel context, 
in which uniform density and magnetic field are assumed on both sides 
of the contact discontinuity and a constant acceleration is considered. However, 
the interface between the synchrotron nebula and the shell of ejecta is 
far from this condition. No analytic theory is available for 
the specific conditions and evolution of the interface. Despite this, 
we will try to derive from the standard theory above information that 
will help us in the interpretation of our numerical results.
We will use the plane parallel criterion (Eq.~\ref{eq:rtch}), with values 
of the various quantities derived from the self similar evolution 
(\cite{chevalier92};\cite{bucciantini04}). 

The first thing to notice is that the above equations can be easily simplified 
in the case one of the two fluids has a much higher density (or enthalpy in 
the relativistic regime) than the other. This is exactly the case in PWN-SNR 
systems. For the shell of ejecta (a non relativistic gas) we can simply consider 
the density, while for the PWN we only need to consider the total pressure.

As shown in previous papers (\cite{bucciantini03}; \cite{bucciantini04}), in 
the framework of the self similar model, the total pressure at the contact 
discontinuity is independent of the wind (or nebular) magnetization, so that
its value and evolution can be derived from a simplified HD description.
It turns out that the shell density is about 3-4 orders of magnitude greater 
that the PWN enthalpy. Therefore the latter can be neglected, 
and Eqs.~\ref{eq:rtch}-\ref{eq:rtlc} simplified.

Assuming a constant pulsar luminosity and ejecta having a density profile
 $\rho_{ej}\propto r^{-\alpha}t^{\alpha-3}$, the PWN expands as $R_{pwn}
\propto t^{\beta}$, with $\beta=(6-\alpha)/(5-\alpha)$. The total pressure 
of the PWN at the contact discontinuity is given by:
\be
P_{tot}(t)=\left[\frac{\beta(\beta-1)}{3-\alpha}+(\beta-1)^{2}\right]\rho_{ej}
(R_{pwn},t)\left(\frac{R_{pwn}}{t}\right)^{2}.
\label{eq:ptot}
\ee
In principle, the magnetic pressure at the boundary can change in time 
as the velocity of the contact discontinuity increases. These variations,
anyway, will be more important in the early phases, and, in general, we 
might assume, to a good approximation, that the magnetic pressure is a 
constant fraction $C$ of the total pressure ($C\simeq 0.5$ at equipartition).

Concerning the radial variations of the enthalpy and magnetic field in the PWN,
we notice that their length-scales are of the same 
order of the nebular radius, so that using the values at the discontinuity 
itself would not be too bad an approximation. On the contrary, the swept up 
shell of ejecta shows a strong gradient of density (J98). To remain within
our simplified approach, we will assume for the shell a uniform density,
equal to the average: 
\be
\rho_{sh}(R_{pwn},t)=\frac{\rho_{ej}(R_{pwn},t)}{(3-\alpha)\Delta}\ ,
\label{eq:rtsh}
\ee
where $\Delta$ is the ratio between the thickness of the shell and the 
nebular radius.
The self similar theory gives a shell thickness corresponding to 
$\Delta \approx 0.02-0.04$ for values of $\alpha$ ranging from 0 to 2 (J98). 
These values of $\Delta$ refer to the adiabatic case, while smaller 
values are found if radiative cooling is considered. 

Another important parameter is the acceleration of the system. The effective 
acceleration of the contact discontinuity can be easily derived in the 
self similar model and it reads:
\be
g=\beta(\beta-1)\left(\frac{R_{pwn}}{t^2}\right)\ .
\label{eq:acc}
\ee

To preserve the overall self-similarity we will deal with constant angular 
sizes of the perturbations. The linear wavelength of the perturbation, $\lambda$,
will increase in time as the nebula expands:
\be
\lambda=\theta R_{pwn}\ ,
\label{eq:lb}
\ee
with $\theta=const$ the angular size.
We define the value of the critical density from Eq.~\ref{eq:rtbc} as:
\be
\rho_c=\frac{B^2}{g \lambda}\ ;
\label{eq:rhoc}
\ee
the system is unstable for $\rho_{sh} > \rho_c$.
In the plane parallel theory the ratio between the shell and critical 
density would determine the stability properties of the system, as a 
function of the size of the initial perturbation. For limitations on the
validity of this approach depending on the values of $\lambda$ see 
Section~2.2. We will follow the 
same approach to derive a stability criterion and to determine which
parameters of the PWN-SNR system are important in this regard.
After some algebra we find from Eqs.~\ref{eq:ptot}-\ref{eq:rhoc}:
\be
\frac{\rho_{sh}}{\rho_c}\simeq\frac{1}{C}\frac{1}{8 \pi \Delta}
\frac{1}{1+(3-\alpha)(\beta-1)/\beta}\;\theta\ .
\label{eq:rtst}
\ee
In the range $\alpha =0-2$, assuming equipartition ($C\sim 0.5$) and no cooling, 
we find that this ratio exceeds unity only for perturbations with
$\theta \simgt \pi/10$.

Let us discuss in more detail Eq.~\ref{eq:rtst}. The first point to notice 
is that the definition of the stability criterion does not have any time 
dependence. This is a direct consequence of the self-similarity assumed 
for the evolution: in this case, perturbations can be stable or unstable, 
but no transition between these two regimes is allowed for a given 
angular size. If self-similarity is a good approximation for the evolution,
the stability of the interface is independent on the specific age of 
the PWN-SNR system. Thus numerical results, that in principle refer to 
a specific moment of the nebular history, can be generalized. In general,
the self-similar evolution of PWNe is determined by three dimensional parameters:
the pulsar luminosity and the energy and mass ejected during the SN. 
Nonetheless, none of these quantities appears in Eq.~\ref{eq:rtst}, where
only a slight dependence on the density profile of the ejecta is present.

Another point that we want to emphasize is that, as shown by Shivamoggi (1982), 
a strong magnetic field might lead to larger values of the critical density. 
A very weak wind magnetization 
is sufficient to make the interface stable, even weaker than what is 
usually assumed, based on the dynamical constraints of the nebular evolution
(\cite{kennel84}). On the other hand, due to the steep
internal gradient, the effective value of the shell density in our problem 
is likely higher than $\rho_{sh}$ (Eq.~\ref{eq:rtsh}).

With regard to the time evolution of the system, as pointed out 
by \cite{bucciantini04}, if the effect of pulsar spin-down is 
included, for times greater than the spin-down characteristic time, 
the expansion tends asymptotically to a linear behavior 
$R_{pwn}\propto t$. This means that the local acceleration will drop to 
zero and the RT instability will be suppressed at later times more 
efficiently than in the constant pulsar luminosity case.

One of the major differences between our problem and the classical description
is that the latter assumes a time independent
background condition and does not take into account dynamical effects 
of the evolution itself. While we do not expect this to have strong effects on the 
stability (which is an instantaneous condition), if the time scale for the 
development of the instability is of the same order the time scale of the 
PWN-SNR system evolution, the growth of the perturbation will be significantly 
altered.

During the expansion, new, more homogeneous material gathers on the shell. 
In this case, instead of an exponential growth, we expect that the perturbation
will increase as a power law in time (i.e. less efficiently).
This is simply seen as follows.
From Eq.~\ref{eq:rtch} the {\it instantaneous} growth rate in the HD limit is:
\be
w(t)=\sqrt{\frac{2 \pi \beta(\beta-1)}{\theta}}t^{-1}\ ,
\label{eq:instg}
\ee
where we have used the expressions for $g$ and $\rho_{sh}$ derived above and the
fact that the density of the ejecta is much higher than the enthalpy of the PWN.
Again, in Eq~\ref{eq:instg}, there is no dependence on the pulsar luminosity, 
nor on the energy and mass involved in the SN. 
Integrating in time, the growth of the perturbation turns out to be
given by:
\be
\exp{\int_{t_0}^{t}\sqrt{\frac{\beta(\beta-1)}{\theta}}z^{-1}dz}\ .
\ee

\subsection{Rayleigh-Taylor and Thin-Shell instability}

The interface between the PWN and the shell of swept up ejecta will be in 
principle subject also to other kinds of instabilities, resulting in a more 
complex evolution. One of the most important types of instability that may 
arise in wind bubble systems is the Thin-Shell (TS hereafter) instability. 
The TS instability has been studied in detail in various environments, from 
the generic bounded slab problem (\cite{vishniac83}; \cite{vishniac94}) to 
radiative SNR (\cite{chevalier95}; \cite{blondin98}) and planetary nebulae 
(\cite{carpenter01}). 
This instability arises when a thin shell (thickness much smaller than 
the curvature radius) is forced to bend as a consequence of density 
perturbation. In the plane parallel case the instability results in the 
formation of a strongly turbulent mixing layer. However, in spherical 
geometry, the shell can bend without undergoing disruption (\cite{carpenter01}).
In our case, the TS instability will thus tend to compete with the RT 
instability for large wavelength perturbations, i.e. perturbations on scales
much larger than the relative thickness of the shell ($\theta >> \pi/50$). 
For smaller scales the RT instability might have the fastest growth.
Numerical simulations might help to determine the threshold between the regimes
of dominance of either kind of instability. 

\section{Numerical simulations}

All the simulations have been performed using the recently developed scheme 
by Del Zanna et al. (\cite{delzanna02}; \cite{delzanna03}). We refer the reader 
to the cited papers for a detailed description of the code, and of the equations 
and algorithms employed. This is a high resolution conservative (shock-capturing) 
code for 3D-RMHD based on accurate third order reconstruction ENO-type algorithms 
and on an approximate Riemann solver flux formula (HLL) which does not make use 
of time-consuming characteristics decomposition. The code employs the 
Upwind Constrained Transport algorithm (UCT) to ensure
the  divergence free constraint on the magnetic field to machine accuracy
(\cite{londrillo00};\cite{londrillo04}). 
The soleinoidal condition is especially important for the present 
application because we are interested in wave modes parallel to the magnetic 
field direction.

In principle one should use a different adiabatic coefficient for the PWN 
($\Gamma=4/3$) and the SN ejecta ($\Gamma=5/3$), however, in numerical 
simulations, the use of two different adiabatic coefficients on a contact 
discontinuity with a very large density jump (density may change by factors 
of order ~$10^{5}-10^{6}$), leads to the formation of spurious waves that 
tend to propagate back into the PWN (\cite{shyue98}; \cite{karni94}; 
\cite{kun98}; \cite{bucciantini03}). Such spurious oscillations may trigger and 
artificially amplify the formation of Kelvin-Helmholtz instability at the 
interface itself, thus affecting the stability properties of the system. This 
is the reason why we have decided to assume a unique adiabatic coefficient,
$\Gamma=4/3$, both for the PWN and the SNR. It should be noticed that this 
choice of adiabatic coefficient allows us to reproduce the correct internal 
structure of the PWN and, as previously discussed (Bucciantini et al. 2003), 
leads to a value of the total pressure at the boundary that is independent 
on the wind magnetization. As a consequence the time-evolution of the 
contact discontinuity and hence the acceleration at the origin of the RT 
instability turns out to be independent on the wind magnetization. As a side 
effect, this allows a direct comparison between the hydro and MHD simulations.

Previous simulations by J98, in the hydrodynamical regime, have been performed 
using a value $\Gamma=5/3$. When compared to those, our simulations lead to
a slightly more compressed shell, but we have verified by means of classical HD 
simulations that there is no significant difference in the growth rates of the 
instability.

For the same reason we have used second order reconstruction with ``monotonized 
centered'' limiter. We have verified that higher order reconstruction, or the
use of more compressive limiters, such as ``superbee'', turns out to enhance 
the numerical noise at the interface, above the level that can be dissipated 
by numerical diffusion. The signature of this effect is an excess of secondary 
KH features. We want to point out that the problems related to the use of 
sharper reconstruction are much more severe in our case than in non-relativistic 
cases, where the density jump is actually reduced by an amount of order the
Lorentz factor of the wind.

\subsection{Initial conditions}

The task of following the evolution of PWN-SNR system from few years after 
the SN explosion up to an age of 2000-3000 years (when the interaction of 
the PWN with the reverse shock in the ejecta is supposed to happen) requires 
a huge computational grid and turns out to be extremely time consuming. 
Moreover, the CFT algorithm employed to guarantee the divergence free condition 
for the magnetic field is not easily implementable together with Lagrangian remapping 
(as, for example, in Blondin et al. 2001; Wolfgang \& Blondin 1997), 
and, on the other hand, implementation of a moving grid is not a trivial task 
in a proper relativistic contest (Lorentz transformations instead of Galilean). 

The reason why these simulations are very time-demanding is that there are
two very different time-scales involved in the problem: the evolution
time-scale for the PWN, which is set by the velocity of the ejecta, and the
integration time-step, which, on the other hand, is determined by the speed
of the wind ($c$). In order to reduce the computational time one may 
artificially 
make these two velocities closer than they would be in reality, taking 
particularly fast ejecta or a slower wind. We reduced the ratio between these 
two speeds by a factor of 10, gaining a factor of 30 in computational time,
although still keeping all the relativistic effects associated with a high 
Lorentz factor wind. 

We have chosen to evolve the system on a 2D spherical uniform grid 
($1500 \times 125$ cells in the r-$\phi$ plane) corresponding to an equatorial 
section, extending in radius from 0.2 ${\rm Ul}$ to 21.5 ${\rm Ul}$. 
Simulations have been 
carried out for different angular sectors from $\pi/32$ to $\pi/6$. The 
radial resolution has been chosen so as to have the shell resolved on at 
least 20 cells. Concerning the number of cells in the angular direction 
we have verified that having 125 cells is a good compromise between 
the need for high resolution and efficiency. At lower resolution numerical 
diffusion is found to damp the formation of secondary KH although the growth 
rate of the RT instability is not affected.
Concerning the unit length ${\rm Ul}$, it may be useful to mention its value in 
physical units for a typical PWN-SNR system. For a pulsar wind luminosity 
(assumed to be constant in time) of $10^{40}$ erg/s and an energy release 
in the SN explosion of $10^{51}$ erg associated with $3M_{\odot}$ of ejecta
one has ${\rm Ul}=1$ ly.
However, we want to emphasize that if the evolution can be approximated as 
self-similar the properties of RT instability (stability criterion and 
growth rate) for a given angular perturbation will be independent of the 
specific choice.

No magnetic field is assumed in the SNR and no radiative cooling is included. 
An ultrarelativistic wind with Lorentz factor $\gamma=100$, $p/\rho c^2=0.01$  
and toroidal magnetic field is injected at the inner radius. Continuous 
conditions ($0^{th}$ order extrapolation) are used at the outer boundary. 
Periodic conditions are imposed in the angular direction.

Initial conditions at time $t_0$ (corresponding for the above described system 
to an age of 200 years after the SN) are derived using a much higher resolution 
1D simulation starting from a time $t \ll t_0$ after the SN explosion. This 
choice allows us to start from a situation in which the system is almost 
completely relaxed on the self-similar solution. No transient phase is observed,
contrary to the results in J98. The evolution is followed up to an age of 
about $10\ t_0$. An initial perturbation is imposed on the swept up shell 
density:
\be
\rho_{sh}=\rho_{sh,unpert}+ \Delta \rho \cos{(2\pi\phi/\theta)}
\ee 
where $\phi$ is the angular coordinate, $\theta$ is the angular scale of the 
perturbation and $\Delta \rho$ is the initial amplitude of the density perturbation.
All the simulations consider just a single wavelength. To cut down computational 
time no multimode case has been studied. No perturbation is assumed in the free 
expanding ejecta.

Different values of the magnetization parameter $\sigma$ (with $\sigma$ the ratio
between magnetic and total luminosity of the wind) have been used. 
In the standard one-dimensional PWN theory 
(\cite{rees74}; \cite{kennel84}) the value of the magnetic field at the contact
discontinuity is determined by the magnetization of the wind and by the 
nebular expansion velocity. Having chosen to increase by a factor of 10 the 
ratio between the latter and the velocity of the wind with respect to a typical 
case, a magnetization correspondingly higher than the usually mentioned values
is required to achieve equipartition at the boundary.

larger than usually mentioned values of $\sigma$ are required to achieve 
equipartition at 
the boundary. In our settings a wind with $\sigma \simeq 0.03$ will result 
in equipartition between magnetic and thermal pressure at the boundary.

Simulations have been carried out also for lower values of $\sigma$ (namely 
$\sigma=$ 0.01, 0.005, 0.0025) which result in a pressure dominated 
interface. Comparison between simulations with different magnetization is aimed
at estimating how the development of the instability is affected by the presence
of a parallel magnetic field, and, in particular, whether in a magnetic dominated 
regime, magnetic compression is sufficient to stabilize the system.

\section{Results}
\subsection{Hydrodynamical Simulations}

\begin{figure*}
\resizebox{\hsize}{!}{\includegraphics{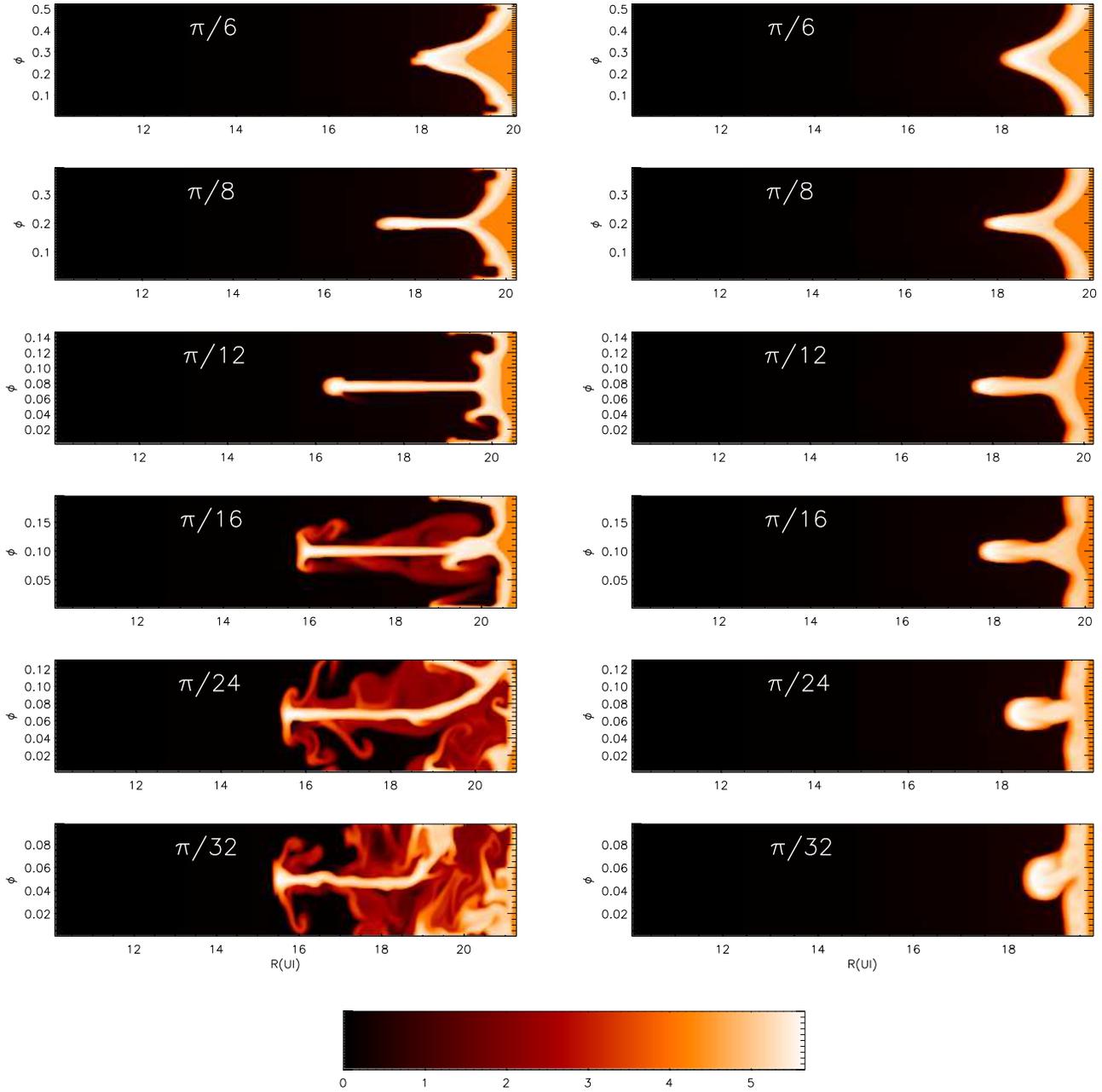}}
\caption{Structure of the RT unstable interface for different angular scales 
of the initial perturbation (logarithmic color-scale plot of the density) at 
$t=10 t_0$. The panels on the left refer to the case $\sigma=0$; those on the 
right to the case $\sigma=0.0025$. Notice that in the magnetized case the 
efficiency of RT is maximum for angular scales of the initial perturbation
between $\pi/12$ and $\pi/16$.}
\label{fig:rt1}
\end{figure*}
\begin{figure*}
\resizebox{\hsize}{!}{\includegraphics{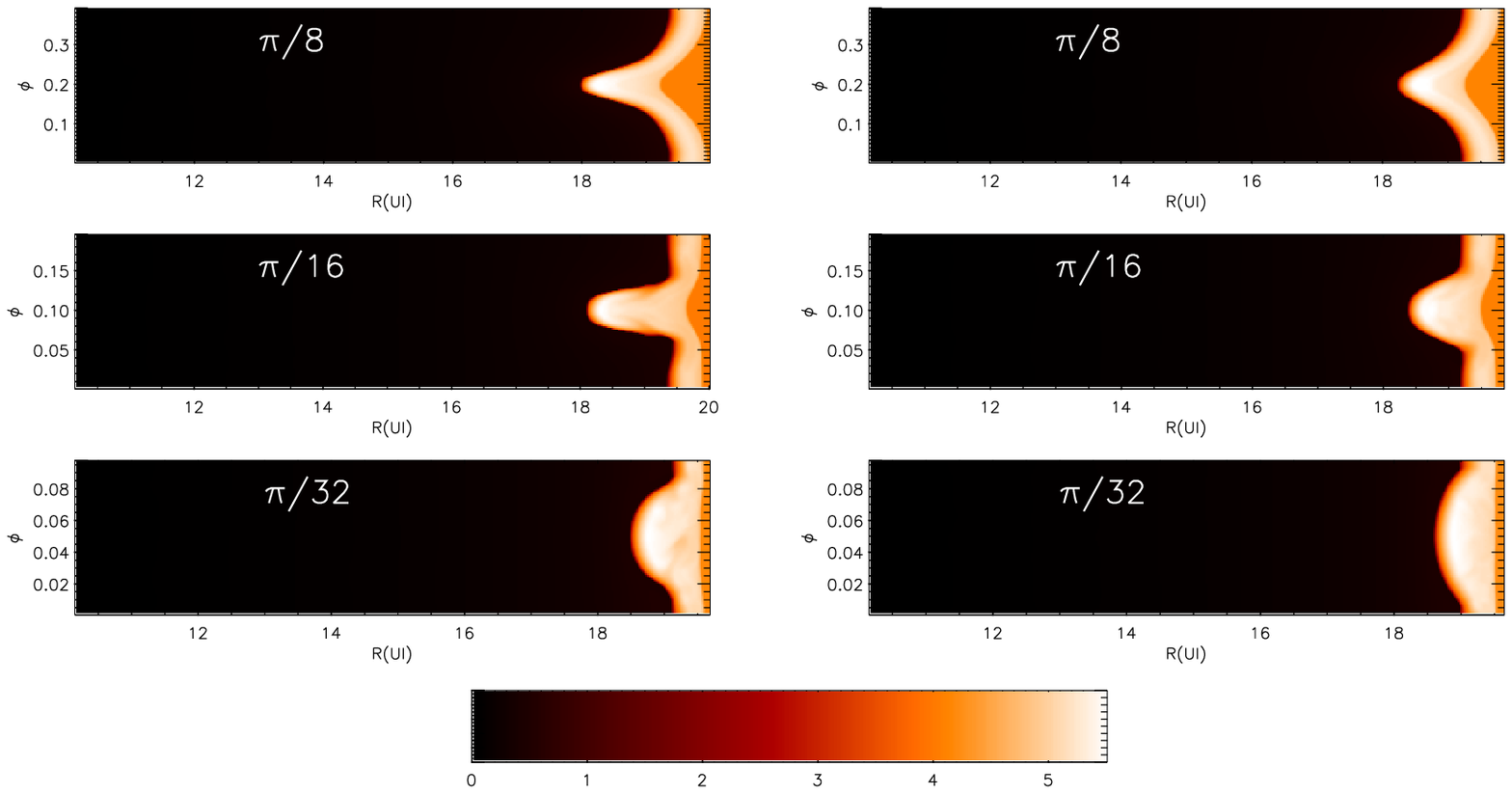}}
\caption{Same as in Fig.~\ref{fig:rt1} but for higher values of the PWN
magnetization. The panels on the left refer to the case $\sigma=0.005$; 
those on the right to the case $\sigma=0.01$. Now the maximum growth rate 
for RT has moved to larger angular sizes $\sim\pi/8$. }
\label{fig:rt2}
\end{figure*}

We start our investigation of the development of RT instability by considering 
the hydrodynamical regime. This is a good approximation when the tension and
pressure of the magnetic field are not important for the dynamics of the
fluid at the interface. Previous work by J98 was aimed at a study of the 
global evolution of the PWN-SNR system under the effect of the multimode
perturbation originating from numerical noise. We will here focus instead 
on the development of monochromatic perturbations of different angular 
extent, analyzing their effects on the shape of the shell in the relativistic
regime. 

The HD simulations where carried out for perturbation of angular sizes: 
$\pi/6$,  $\pi/8$,  $\pi/12$, $\pi/16$, $\pi/24$, $\pi/32$. In all the 
simulations whose results are here presented the amplitude of the initial 
density perturbation was taken to be $\Delta \rho/\rho=0.25$. The choice
of such a large amplitude of the initial perturbation was motivated by 
computational reasons (we wanted the finger to extend on a large number of 
cells of the fixed computational domain in a reasonable time). However 
we checked that the adopted value of $\Delta \rho/\rho$ was still small
enough not to affect the growth-rate of the fingers (we have checked 
that starting with a perturbation 0.1 we end up with a finger that is 
about one half shorter).

Looking at Fig.~\ref{fig:rt3}, initially we see a superlinear phase 
(see also \cite{jun95}). The large amplitude assumed for the perturbation, 
together with the fact that the initial unperturbed state of the system was 
derived from a much finer grid, are at the origin of this transient, lasting 
for a time $t \simlt t_0$. At later times, the interface enters a linear 
growth phase, during which it bends to form a bump into the lighter fluid. 
Finally a non-linear phase begins, when secondary KH-type 
instabilities at the head of the finger result in the formation of a 
mushroom cap.

In Fig.~\ref{fig:rt1} we can notice a number of differences in the simulation 
results corresponding to initial perturbations of different angular size.  
Large scale perturbations ($\pi/6$,  $\pi/8$) lead to a uniform distortion of 
the whole shell. This behavior is more typical of the TS rather than RT 
instability. When the former is the main process at work, the shell material 
is not efficiently advected toward the bump, and the density contrast between 
the wings and the center of the interface distortion keeps almost constant during 
the evolution.

\begin{figure}
\resizebox{\hsize}{!}{\includegraphics{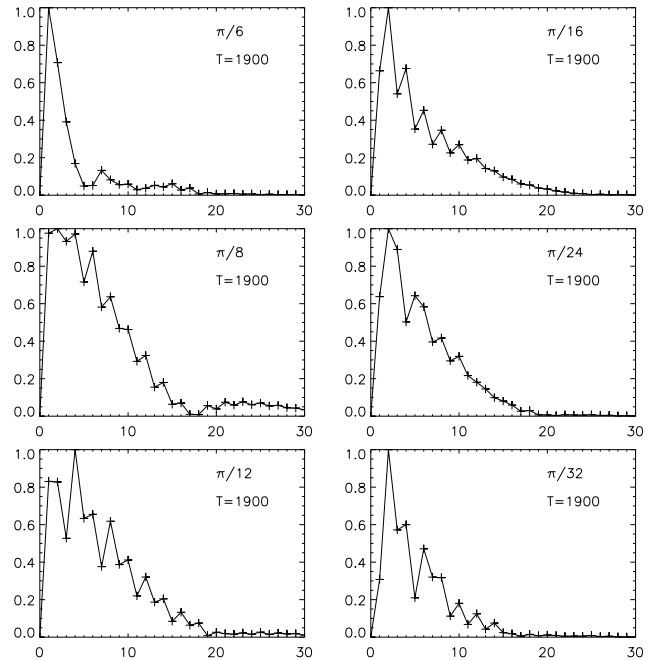}}
\caption{Spatial Fourier transform of the column density $\eta$
in the region between 
the head of the finger and the shell of swept-up ejecta for the case $\sigma=0$. 
The transform is taken at a time $t=10 t_0$ and the average has been subtracted.
The different panels correspond to different angular sizes $\theta$ of the 
initial perturbation. The horizontal scale is in units of $1/\theta$. Except
for the case $\theta=\pi/6$ we observe a cascade to small scales associated 
with the formation of a narrow finger and structure associated to KH instability.}
\label{fig:rt5}
\end{figure}

There is a transition between the regime of dominance of RT versus TS
instability at wavelengths corresponding to $\theta \sim\pi/8$. For this value 
of $\theta$ the whole shell still shows a bending typical of the TS but a 
rather elongated finger, likely related to an intervening contribution of RT 
instability, is now present at its bottom. This behavior can also be 
appreciated looking at Fig.~\ref{fig:rt5} where the power spectrum of the 
normalized density distribution is shown. For $\theta=\pi/8$ we see that the 
power is still on the scale of the original perturbation, while we start to 
observe a decay to smaller scales, associated to the formation of 
a narrow finger only for $\theta \simgt\pi/12$.

As expected the growth of the finger is more rapid at larger wave numbers
and it seems to converge for small scale perturbations. This is probably the 
effect of the intervening KH instability that drags away material from the
head of the finger: this effect naturally occurs earlier at smaller scales.

Another point to notice is that no secondary KH-type instability along the 
finger is actually present in simulations with $\theta >\pi/16$. This is due 
to the combination of two main effects: first of all, the larger growth rate
of RT instability at smaller wavelengths causes a larger shear along the 
boundaries of the finger providing a more effective trigger for the 
KH instability; moreover, as the resolution in angle improves, numerical noise, 
that may act as a seed for the instability is dissipated less efficiently. 

Our simulations show that the formation of mushroom caps on top of the fingers
is a quasi-periodic process. When a mushroom cap is formed its typical density
is about 2 orders of magnitude below that in the head of the finger, and the 
drag exerted on its wing by the lighter fluid can force it to shift toward the
base  where it collides with the swept up shell of ejecta. In the meanwhile, 
shear on the head causes the formation of a new mushroom cap and the process 
starts again. We notice also that in none of our cases the finger penetrates 
the free flowing wind region and the thickness of the mixing layer extends in
the case $\theta = \pi/32$  for about one quarter of the size of the entire
nebula. As the head of the finger recedes toward the termination shock, the
velocity of the relativistic fluid in front of it increases and so do the 
drag and the shear thus reducing the growth itself. This result is in 
principle related to the initial 
perturbations. However the perturbation we adopted is fairly high, and even 
extrapolating to later times in the evolution it would take $\sim40t_0$ to penetrate
 the termination shock. Given that the free expansion phase lasts for about
3000-4000 years this means that, even neglecting the high shear the head of 
the finger would be subject when moving to the inner part of the nebula, the 
instability needs to start since ~50 yr after the SN in order to 
affect the relativistic wind at very late times (spin-down 
will cause the termination shock to recede even more toward the pulsar, 
Bucciantini et al. (2004)), and it is not clear if at such an early time the
PWN has already set into self-similar expansion. So we deem 
this result can be generalized as a common property of the filamentary 
network in PWNe.

The efficiency with which material is actually forced to converge into the
finger strongly depends on the size of the initial perturbation. The less
the mass that moves from the shell into the finger, the less the shell 
deformation will be: in the case of shortest wavelengths both the surface 
density and the total mass
in the shell are actually less than in the unperturbed case, causing a slightly 
larger nebular radius. Looking at Table~\ref{tab:tab1} we see 
that in the HD case the mass in the mixing layer 
(finger plus turbulent KH structure) ranges from 2 to 4 times  
the mass in the shell: a result in agreement with previous simulations in 
classical HD regime by J98. However we want to stress that even a large
change in the total mass of the shell implies only minor changes in the
evolution of the PWN size. In the case when $\theta = \pi/32$, despite 
80\% of the total swept-up mass is in the mixing layer (that, as we 
mentioned, extends for about one quarter of the size of the nebula) the final 
radius of the PWN is only 10\% larger.

From the point of view of observations, it may be interesting to notice the
following. When fingers are well developed, the density in the head can be
2-3 times higher than the maximum value reached in the shell. Therefore, if
one expects radiative cooling to be effective in the shell, it will be much
more so in these clumps, where the gas may be only partially ionized.

Looking at the fingers' evolution at different $\theta$, we can see that 
as the KH turbulent layer becomes more developed, the fingers undergo
more significant deformation. For $\theta=\pi/16$ the finger is still
approximately straight and attached to the shell at its basis. 
When secondary turbulence along the 
finger becomes more important, as in the case $\theta=\pi/32$, it may cause
the finger to bend and eventually detach from the shell. The outcome of this
process is a structure formed of dense clumps embedded in a turbulent mixing 
layer. Trace of the disruption of the finger is present in Fig.~\ref{fig:rt5}:
The tail at small wavelengths is more developed in the cases $\theta=\pi/16,\pi/24$
than for  $\theta=\pi/32$, because in the latter the finger have bent and is
 undergoing disruption while in the former cases the finger even if shorter
 is still straight and attached to the shell.

As noticed by Wang \& Nepveu (1983) another interesting aspect of RT 
instability is the complementary behavior of the heavy and light fluids: 
``as one falls in a given sector of the boundary, inducing the other to 
rise in other sectors of it, large scale vortex motions are set up which 
result in the formation of interlocking mushroom structures''. This is 
exactly what we observe at the boundary of the simulation box in $\phi$. 
Large circulatory motions produced by the development of the primary RT 
fingers at the center of the box, coupled with secondary KH instability
give rise to the formation of smaller fingers at the boundary itself. 
These fingers are present even in simulations carried out with a lower 
resolution, where mushroom cap formation is damped by numerical diffusion.

Looking at the turbulent structure in the mixing layer we observe a rapid 
growth of the curl of the velocity field once the secondary KH starts to 
create caps and vortexes along the finger. Once the mixing layer is well 
developed the vorticity seems to saturate and its growth is strongly reduced.

In all cases where RT is the dominant instability ($\theta \ge \pi/8$), 
the thickness of the finger progressively decreases until it is only of 
order one tenth of the size of the initial perturbation. This can be largely 
attributed to the ram pressure exerted by the surrounding relativistic fluid, 
as it moves around in a circulatory pattern. Again this can be seen looking 
at Fig.~\ref{fig:rt5} where the width of the power spectrum of the density
distribution is $\sim 10$ in unity of the wave number of the original 
perturbation.

As far as the power spectrum of the mass distribution is concerned, we found it
useful to introduce the radial column density $\eta$ as the quantity to 
look at. We defined $\eta$ as:
\be
\eta(\phi)=\int_{\cal L} r^2 \rho(r,\phi) dr
\label{eq:eta}
\ee
where $\cal L$ is the radial extent of the mixing layer (from the front shock
in the swept up shell to the head of the finger).
In Fig.~\ref{fig:rt5} we show the power spectrum of $\eta(\phi)$. We see from 
the figure that, except for the case $\theta=\pi/6$ where almost all the power 
is still at the scale of the original perturbation, in all other cases there is
evidence for the formation of structures on scales as small as 
$\sim 1/20 \theta$. 

We want to point out that we decided to trace the growth of the fingers by 
measuring their length. With respect to Jun et al. (1995), where the growth rate 
is measured by using the amplitude of the density perturbation, one might expect 
the comparison not to be straightforward. Our choice was motivated by the need to
 provide a growth rate by means of parameters (the length of the finger) that can 
in principle be easily
observed, while Jun et al. (1995) choice was probably dictated by the need of
a comparison with theory (in our case there is no theory for comparison).
However if one follow the evolution of the density perturbation (Table~\ref{tab:tab1})
we find that the standard behaviors of Fig.~\ref{fig:rt3} (like saturation 
and magnetic compression) is retrieved.

\subsection{Magnetohydrodynamical Simulations}

\begin{figure}[h!!!!!!!]
\resizebox{\hsize}{!}{\includegraphics{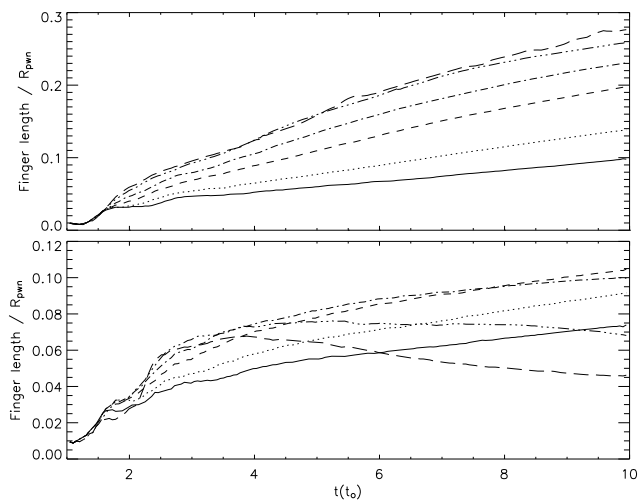}}
\caption{Evolution of the ratio between the length of the finger 
(radial extent of the mixing layer in the turbulent HD case) and the 
radius of the PWN $R_{cd}$. The various curves refer to different 
initial perturbations: solid $\theta=\pi/6$; dotted $\theta=\pi/8$; 
dashed $\theta=\pi/12$; dot-dashed $\theta=\pi/16$; 
dash-triple-dot $\theta=\pi/24$; long-dashed $\theta=\pi/32$. 
The upper panel refers to the case $\sigma=0$; the bottom panel to 
$\sigma=0.0025$. After a superlinear phase, lasting for about $0.5 t_0$
for all values of $\theta$, we observe a linear phase that for small
$\theta$ in the HD regime last up to about $t\sim6t_0$, while in the MHD cases
it stops earlier at around $t=\sim4t_0$.
The growth of the finger saturates at later times for the smallest wavelengths.
See text for a detailed description.}
\label{fig:rt3}
\end{figure}
\begin{figure}[h!!!!!!!]
\resizebox{\hsize}{!}{\includegraphics{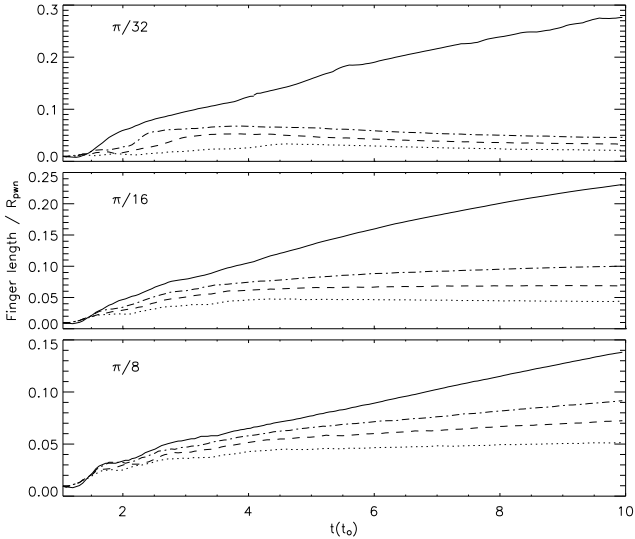}}
\caption{Evolution of the ratio between the length of the finger 
and the radius of the PWN $R_{cd}$. The upper panel refers to 
$\theta=\pi/32$; the middle panel to $\theta=\pi/16$; the bottom one
to $\theta=\pi/8$. Different curves refer to different values of the
wind magnetization: solid $\sigma=0$; dotted $\sigma=0.001$; 
dot-dashed $\sigma=0.0025$.}
\label{fig:rt4}
\end{figure}

The widely accepted picture for PWNe leads to believe that at the contact 
discontinuity with the SNR the relativistic bubble should be magnetically 
dominated. Both the one dimensional models by Kennel \& Coroniti (1984), 
and Emmering \& Chevalier (1987), as well as the 2D stationary model by 
Begelmann \& Li (1992), give a ratio between the magnetic and the thermal 
pressure at the boundary of the nebula close or above equipartition. To 
investigate how the magnetic field affects the development of the RT 
instability in a representative regime, the pulsar wind magnetization has 
to be chosen so as to guarantee the proper conditions at the interface. 
Given our velocity normalization, as well as the PWN-SNR evolution, we 
have determined that a pulsar wind with $\sigma=0.03$ results in an 
equipartition condition at the boundary.

It is well known that the presence of a strong magnetic field can suppress the
RT instability. In fact, for $\sigma=0.03$ no finger is formed for any of the
initial scales of the perturbations considered: only a negligible deformation
of the contact discontinuity is present, in spite of the large density
perturbation amplitude (25\%) adopted. 

To better understand how strong the suppression of RT instability could be, 
and what modifications are to be expected when the system is unstable, we have 
also investigated cases of lower wind magnetization:  
$\sigma=0.01, 0.005, 0.0025$. This will also clarify the transition between 
stable and unstable regimes. 

The $\sigma=0.01, 0.005, 0.0025$ cases give a PWN where the ratio between 
magnetic and thermal pressure at the contact discontinuity is, respectively,
$P_B/P_t \sim 1/5, 1/15, 1/30$. The system is progressively more stable
with increasing magnetization (Figs.~\ref{fig:rt1}-\ref{fig:rt3}), but even 
a weak magnetization has important effects on the finger structure and evolution.

\begin{figure}
\resizebox{\hsize}{!}{\includegraphics{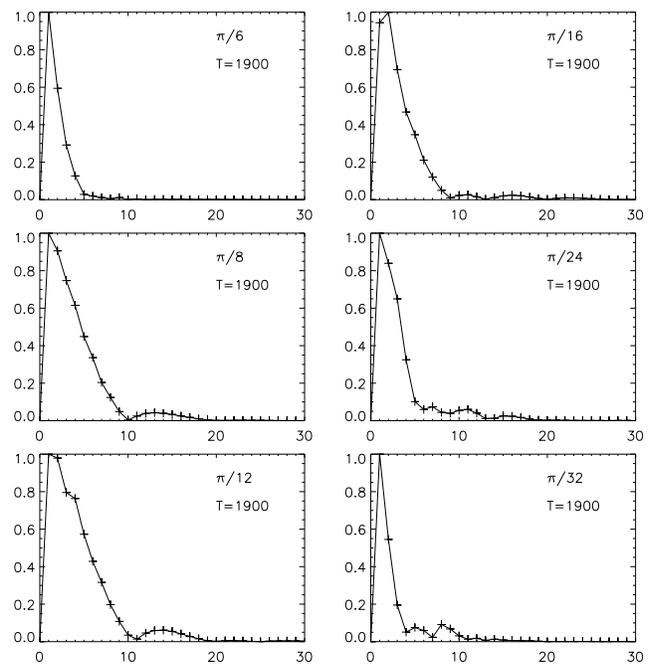}}
\caption{Same as Fig.~\ref{fig:rt5} but for the case $\sigma=0.0025$. 
We notice that now a bump appears in the power spectrum in most panels at 
around $1/10-1/20\ \theta$. This is associated to the formation of a 
thick finger.}
\label{fig:rt6}
\end{figure}

A comparison between the case with $\sigma=0.0025$ and the HD simulations 
shows that:
\begin{itemize}
\item The growth rate is strongly reduced. For all the magnetization values we
considered, the growth of the finger is maximum for 
$\theta$ in the range $\pi/12-\pi/16$ (Figs.~\ref{fig:rt1} and \ref{fig:rt3}). 

\item Secondary KH instability is completely suppressed. No mushroom cap, nor 
secondary fingers at the boundary are formed, and the density profile of
the shell is not much affected. In the magnetic case there is no turbulent
mixing layer and the CD is well defined. There is no increase 
in the curl of the velocity field, at variance with the correspondent 
turbulent HD cases. The flow is still laminar.

\item The fingers tend to be thicker and do not show fragmentation. 
The cascade to small scales is much less efficient than in the HD case, as 
it is clear from the power spectrum of $\eta$ (Fig.~\ref{fig:rt6}). 
The density in the head of the finger is about a factor 2-4 times higher than
the maximum in the shell. This ratio tends to be slightly higher than in the 
corresponding HD case: this is related with the fact that now the finger is
more compact (see also the presence of a secondary bump in the power spectrum
in Fig.~\ref{fig:rt6}).

\item There is a maximum length the finger can achieve, that depends on the 
wavelength and on the magnetization. We found that once the magnetic field 
compressed in front of the finger reaches values close to equipartition, 
the growth stops and in some cases the length of the finger is even slightly 
reduced (Fig.~\ref{fig:rt3}).

\item In all cases when a finger is formed, the magnetic field lines are
bent. Depending on $\sigma$, the radial component of the field that arises
can become comparable (in the $\sigma=0.01$) 
and/or greater ($\sigma=0.0025$) than the toroidal component.
\end{itemize}

As the magnetization increases, for small values of $\theta$ 
(typically $\theta=\pi/32$ in our simulations) a bump of angular size 
close to that of the original perturbation is formed, rather than a narrow
finger. The evolution of the nebular radius is not much affected because
not enough mass is dragged into the finger (see Table~\ref{tab:tab1}) and 
the shell thickness is almost constant during the evolution.
Looking at Fig.~\ref{fig:rt6} we can seen the same behavior from the 
density distribution. Again we notice that a tail to small scales is
developed only for the cases $\theta=\pi/8-\pi/16$, where the growth of the finger
 has not stopped yet (Fig.~\ref{fig:rt3}), while it is suppressed at smaller scales
due to magnetic compression. We also notice that the finger tend to be thicker
than in the analogous HD case. Despite the morphological difference between
the cases $\theta=\pi/32$ and $\theta=\pi/6$ the density distribution is fairly 
similar apart for the presence of secondary bumps that are indicative of the
presence of a finger with a flat density distribution.
Simulations are able to reproduce the density gradient inside the swept-up
shell of ejecta and the local density at the interface is about one order
of magnitude higher than the average value that we use in
Eq.~\ref{eq:rtst}. In principle one might expect the effective
density to be higher at smaller scales (where the average should be computed 
on a scale of order $\lambda$ rather than on the entire thickness of the shell),
hence satisfying more easily the requirements for instability. 
Despite this, when a magnetic field at equipartition is present,
even in the case $\theta=\pi/6$, which in principle should be below
the threshold value ($\theta\sim\pi/10$) for stability provided by Eq.~\ref{eq:rtst},
 no deformation is observed (neither a relaxation to the imposed initial 
perturbation). Given the simplified 
assumptions used to derive the threshold value itself, like constant density or 
steady state background, and the reduction to 2 dimensions
 of our numerical simulations we are not able to specify the reason 
for such results.
It well might be an effect of a strong magnetic field, as anticipated is 
Sect.~\ref{sect:stcrit}. Introducing the correction provided by Shivamoggi (1982)
the threshold value should decrease to  $\theta\sim\pi/5$, 
in agreement with the numerical results. In the case of a magnetic field five 
times below the equipartition value, we find that a fingers forms for all the 
values of $\theta$ we investigated, in agreement with prediction from Eq.\ref{eq:rtst}.
however even in the cases where a 
finger develops its growth seems to saturate when the field in the head is 
compressed to equipartition value thus suggesting the presence of some 
stabilizing effect that the standard theory does not account for 
(further investigation is under development). 
It is interesting to notice that even if the interface does not bent at all,
the density perturbation in the swept-up shell does not disappear: the relative 
amplitude is obviously reduced as a consequence of new material gathering on the
 shell, but the original perturbation is frozen. This confirm that stability does 
not change  with time as the evolution proceeds.
This result shows the limitations of former simulations in pressure dominated 
regimes where the standard stability criterion was found to give good results. 
Nonetheless, other results concerning the shape of the perturbed interface, the 
formation of secondary KH instability, as well as the existence of a maximum 
deformation, still hold (\cite{jun95}).

Our results seem to agree with those of Wang \& Nepveu (1983) in the context of 
accretion onto a compact object. The actual threshold values derived from 
Eq.~\ref{eq:rtst} seems to be underestimated, but further analysis to clarify
this point is required.
 
\subsection{Implications of the MHD results}


Our results show that once the magnetic field is included in the evolution of 
the PWN inside its SNR the statement that the accelerated shell of ejecta 
should be RT unstable is not so trivial as it has been considered. As we 
have already stressed, since the acceleration is basically independent of 
the exact value of the magnetic field, but depends only on the pulsar total 
luminosity and SNR properties, a magnetic field above or close to equipartition 
can suppress the growth of the perturbations parallel to its direction, 
even of the large scale ones.

This has important consequences on the correct understanding of the physical 
conditions both at the boundary of the PWN and inside the shell of ejecta. In 
the case of the Crab Nebula, H96 concluded that the shell must be subject to  
efficient cooling. Using the standard values (\cite{chandrasekhar61}) for the 
critical threshold, H96 finds that the shell must have a density at least one 
order of magnitude greater than the upper limit estimated from the
non-detection of the halo (\cite{sankrit97}) assuming adiabaticity.
However, the radiative cooling hypothesis is not without problems. Even though 
the metallicity could in principle be high enough to give the required cooling 
efficiency, the effectiveness of the cooling process could be partially 
reduced by the presence of a synchrotron emitting nebula, which can provide 
enough UV radiation to keep the material inside the shell from cooling. 
Moreover efficient cooling means that the relative thickness of the 
shell should be reduced by the same factor by which the density is increased. 
A much thinner shell would be subject to more efficient TS instability and the 
threshold between TS and RT would eventually move to smaller scales. As we 
noticed, in our simulations this threshold falls in the range 
$\theta \sim \pi/6-\pi/8$. If the cooling reduces the thickness by a factor 
of 10 we might expect the threshold angular size to drop by about 
the same factor. This means that the observed size (H96) of the perturbations 
in the Crab Nebula would be close to this threshold. 

If cooling is required as a key ingredient for instability in PWN-SNR systems, 
it is highly probable that the efficiency of finger formation was higher during 
the early stages of the evolution. The cooling efficiency, in fact, strongly 
depends on the density of the shell, which is expected to have been much higher 
in the early phases. The system could have experienced a radiative cooling phase 
at the beginning leading to the development of RT fingers. In the following 
evolutionary phases, it could have become non-radiative, due to the decrease
in the shell density. In the case of the Crab Nebula, for example, the general 
consensus is that this young system is close to the above mentioned transition, 
although there is evidence that cooling might still be going on 
(\cite{graham90}).

Another possible explanation for the origin of the filamentary network is that 
it is caused by dense clumps formed during the SN explosion. Clumps are 
currently observed in the remnant of SN1987A (\cite{wang96}). 
In this scenario, rather than to a local interface instability, the fingers 
would be due to the inertia (or buoyancy) of these dense clumps, which have 
resisted the PWN expansion. However our lack of knowledge of the typical SNR 
clumpiness as well as our ability to measure the density in the fingers do not 
allow us to confirm or to rule out this hypothesis.

We must point out that the standard picture of a PWN with magnetic 
pressure close to equipartition at the boundary comes from 
theoretical models that assume the plasma flow to be laminar from the 
termination shock to the contact discontinuity. More recent developments
in this field suggest that the flow structure may be rather complex and,
even for wind magnetizations such that the average nebular field is around
equipartition, the ratio between magnetic and thermal pressure at the
boundary may strongly depend on latitude (\cite{bogovalov02}; \cite{lyubarsky02}; 
\cite{komissarov03}; \cite{delzanna04}) and be below equipartition in
some regions.

\subsection{Limits of the model}

In this section we want to discuss some limitations intrinsic of our approach. 
Apart from the absence of cooling that could provide a key ingredient in the 
correct modeling of the proper physical conditions at the interface, the 
major limitation is the reduction of degrees of freedom of the system with 
respect to a proper 3D case. Unfortunately, proper 3D simulations are 
prohibitive in terms of computational time, even if simplified plane 
parallel cases in magnetic dominated environments surely deserve more attention.
In addition to all of this, we would also like to remind the reader that our
investigation was performed only for single wavelength modes, thus ignoring
all the possible consequences of mode coupling.

In 2D the magnetic field in the head of the finger can be compressed very 
efficiently leading to high magnetic pressure and damping the growth of the 
instability. In a more realistic 3D situation the magnetic field lines are also 
allowed to move in the transverse direction and to slide around the finger. This 
will eventually reduce the compression in the head so that well 
formed fingers can actually protrude much further in the relativistic bubble. 
This effect will also lead to a deformation of the fingers into sheet like 
structures.

\begin{table*}
  \caption[]{Ratio between the mass in the finger and the mass in the shell at T=$5 t_0$
and T=$10 t_0$.}
  $$
 \begin{array}{p{0.1\linewidth}p{0.1\linewidth}p{0.1\linewidth}p{0.1\linewidth}p{0.1\linewidth}p{0.1\linewidth}l}
   T= $5 t_0$& \\
    \hline
    \noalign{\smallskip}
    $\sigma$   &  $\pi/6$  &  $\pi/8$  &  $\pi/12$  &  $\pi/16$  &  $\pi/24$  &  \pi/32 \\
    \noalign{\smallskip}
    \hline
    \noalign{\smallskip}
    0.0    & 0.10 & 0.25 & 0.60 & 0.70 & 1.30 & 2.00 \\
    0.0025 & 0.05 & 0.15 & 0.45 & 0.60 & 0.75 & 0.75 \\
    0.005  & ---  & 0.10 & ---  & 0.50 & ---  & 0.60 \\
    0.01   & ---  & 0.05 & ---  & 0.30 & ---  & 0.35 \\
    0.03   & ---  & 0.00 & ---  & 0.00 & ---  & $---$ \\
  \end{array}
  $$
  $$
  \begin{array}{p{0.1\linewidth}p{0.1\linewidth}p{0.1\linewidth}p{0.1\linewidth}p{0.1\linewidth}p{0.1\linewidth}l}
    T=$10 t_0$ & \\
    \hline
    \noalign{\smallskip}
    $\sigma$   &  $\pi/6$  &  $\pi/8$  &  $\pi/12$  &  $\pi/16$  &  $\pi/24$  &  \pi/32 \\
    \noalign{\smallskip}
    \hline
    \noalign{\smallskip}
    0.0    & 0.30 & 0.55 & 1.20 & 1.80 & 3.50 & 4.20 \\
    0.0025 & 0.25 & 0.40 & 0.70 & 0.75 & 0.80 & 0.75 \\
    0.005  & ---  & 0.40 & ---  & 0.75 & ---  & 0.70 \\
    0.01   & ---  & 0.20 & ---  & 0.60 & ---  & 0.35 \\
    0.03   & ---  & 0.00 & ---  & 0.00 & ---  & $---$ \\
  \end{array}
  $$
\label{tab:tab1}
\end{table*}

\section{Conclusion}

In this paper, for the fist time, we present simulations of RT instability for PWNe 
expanding into the SNR ejecta, including the effect of a magnetic field 
at the boundary. Simulations were carried out for various initial angular 
perturbations. In the HD regime our simulations confirm previous 
results by J98 concerning the nebular size and fraction of mass contained 
in the mixing layer. The higher resolution we adopted allows us to follow the 
detailed evolution of the finger structure. We find that finger fragmentation 
is likely to happen only once a well developed turbulent mixing layer has formed, 
and that, for large scale perturbations, TS instability seems to dominate over 
RT, leading to an overall deformation of the swept-up shell of ejecta. We were 
able to observe the dragging exerted by secondary KH and the formation of 
mushroom caps. From our simulations it seems that fingers do not grow enough
to penetrate the relativistic wind region and the mixing layer extends to about 
one quarter of the nebular radius.

The introduction of a magnetic field turns out to be a key ingredient for 
a correct understanding and modeling of the RT instability. We have shown 
that the stability criterion adapted to the self-similar PWN-SNR evolution, 
does not depend on pulsar wind luminosity, SN mass and energy and does not 
change during the nebula evolution. The result is that the ratio between the 
shell density and the critical density is close to unity if a magnetic field 
around equipartition is assumed at the PWN boundary. 
This result is confirmed by our simulations that show 
that magnetic field close or above equipartition can completely suppress the 
RT instability even for large scale perturbations. A weaker magnetic field is 
able to reduce the growth of the finger leading to a round rather than elongated 
protuberance attached to the ejecta shell. A non-negligible magnetic field can
suppress the secondary KH completely: no turbulent mixing layer is formed, and 
fingers are thicker than in the HD case. This is the main difference with 
respect to former simulations in the pressure dominated regime, where the 
standard stability criterion was found to give the correct result.

As shown by H96 an efficient cooling is required to explain the RT filamentary 
network observed in the Crab Nebula. Our simulations however point out that 
this requirement is even stronger than previously thought, thus favoring the
hypothesis that the fingers formed during the early phases of the system 
evolution, when the density of the shell was higher. 
An alternative to efficient cooling would be the presence of dense clumps 
formed during the SN that may resist the expansion of the nebula. 

Recent work on axisymmetric pulsar winds seems also to suggest a third possible 
explanation for the formation of RT fingers and possibly their latitude 
distribution. If turbulent large scale convective cells are formed, as a 
consequence of ram pressure gradient in the wind, there might be regions at 
the boundary where the magnetic field is below equipartition and the RT 
instability criterion is satisfied. This scenario deserves more attention, 
however 3D global simulations with high resolution are still too much demanding 
in terms of computational time.

\begin{acknowledgements}
This work has been partially supported by the Italian Ministry of University
and Research (MIUR) under grant COFIN2002, by INAF under grant COFIN 
2002, and partly by a SciDAC grant 
from the US Department of Energy High Energy and Nuclear Physics Program. We thank
the North Carolina Supercomputing Centre and Oak Ridge National Lab for their 
generous support of computational resources.

\end{acknowledgements}


\end{document}